\newcommand{\CLASSINPUTbaselinestretch}{1.8}
\documentclass[onecolumn, final, 12pt]{IEEEtran}
\usepackage{amsmath,cases}
\usepackage{amsthm}
\usepackage{amssymb}
\usepackage{cite}
\usepackage{amsfonts}
\usepackage{enumerate}
\usepackage{url}
\usepackage[final]{graphicx} 
\usepackage[mathscr]{euscript} 
\newtheorem{lem}{Lemma}
\newtheorem{defn}{Definition}
\newtheorem{theorem}{Theorem}

\newtheorem{corollary}{Corollary}

\newcommand{\vect}[1]{\ensuremath{\mathbf{\lowercase{#1}}}} 
\newcommand{\rvect}[1]{\ensuremath{\mathit{\mathbf{\lowercase{#1}}}}} 
\newcommand{\mat}[1]{\ensuremath{\mathbf{\uppercase{#1}}}} 
\newcommand{\set}[1]{\ensuremath{\mathscr{\uppercase{#1}}}} 

\newcommand{\re}{\set{R}}
\newcommand{\N}{\set{N}}
\newcommand{\vectornorm}[1]{\left|\left|#1\right|\right|}
\newcommand{\E}[2]{\ensuremath{\mathbb{E}_{#1}\left[ #2 \right]}}

\newcommand{\Omegai}{\set{K}_{i}}
\newcommand{\Omegaio}{\set{K}_{i}}

\newcommand{\Fj}[1]{\set{F}\left(\Omegaio \right)}

\newcommand{\phii}[1]{\phi_{_{#1}}}

\newcommand{\fhi}[1]{\overline{\phi}_{#1,f,q,i}}

\newcommand{\sheta}[1]{\underline{\theta}_{#1,f,q,i,l}}
\newcommand{\Sheta}[1]{\overline{\theta}_{#1,f,q,i,l}}
\newcommand{\rbar}{\overline{r}_{f,q,i}(\boldsymbol{\phi})}

\newcommand{\fcm}{f_{{\rm cm}}(\rho)}
\newcommand{\ftildei}{\tilde{f}_{{\rm cm},i}(\rho)}

\newcommand{\dsum}{\sum\limits_{f=1}^{F_{i}} \sum\limits_{q=1}^{Q_{f,i}}}
\newcommand{\J}{J_{f,q,i}(\rho)}

\newcommand{\orderl}[1]{\leq_{\rm #1}}

\newcommand{\norderl}[1]{\nleq_{\rm #1}}

\newcommand{\pdf}[2]{\ensuremath{f_{#1}\left(#2 \right)}}

\newcommand{\Pe}[2]{\ensuremath{P_{\rm e}^{\rm #1}\left(#2 \right)}} 
\newcommand{\Pei}{\ensuremath{P_{{\rm e,}i}\left(\rho \right)}}

\newcommand{\AvgPe}{\ensuremath{P_{\rm {e}}\left(\rho \right)}}
\newcommand{\AvgPedd}{\ensuremath{P^{''}_{\rm {e}}\left(\rho \right)}}

\newcommand{\Q}[2]{\ensuremath{\mathcal{Q}^{#1}\left( #2 \right)}}

\newcommand{\sphi}{s(\boldsymbol{\phi})}

\newcommand{\iid}{i.i.d}
\newcommand{\F}{\set{G}}
\newcommand{\D}{\mathtt{d}}

\newcommand{\LT}[3]{ \E{#1}{\exp(-#2 #3)}}

\newcommand{\cm}{c.m.}

\newcommand{\redN}{\ensuremath{N^{*}}}

\begin{document}

\title{A New Representation for the Symbol Error Rate}
\author{Adithya Rajan, Cihan Tepedelenlio\u{g}lu, \emph{Member, IEEE} \thanks{The authors are with the School of Electrical, Computer, and Energy Engineering, Arizona
State University, Tempe, AZ 85287, USA. (Email:
\{arajan2, cihan\}@asu.edu). Parts of this work have been submitted to the Asilomar Conference on Signals, Systems, and Computers, 2012, and \cite{paper:AdithyaISIT12}. This work was supported in part by the National Science Foundation under Grant CCF 1117041.} } 
\maketitle

\begin{abstract}
The symbol error rate of the minimum distance detector for an arbitrary multi-dimensional constellation impaired by additive white Gaussian noise is characterized as the product of a completely monotone function with a non-negative power of the signal to noise ratio. This representation is also shown to apply to cases when the impairing noise is compound Gaussian. Using this general result, it is proved that the symbol error rate is completely monotone if the rank of its constellation matrix is either one or two. Further, a necessary and sufficient condition for the complete monotonicity of the symbol error rate of a constellation of any dimension is also obtained. Applications to stochastic ordering of wireless system performance are also discussed.

\end{abstract}

\section{Introduction}
\label{sec:Intro}
An important performance metric in digital communications is the symbol error rate (SER). The convexity of the SER in the signal to noise ratio (SNR) plays a critical role in various optimization problems \cite{paper:azizoglu96,paper:lin03}. Convex SERs have a negative first derivative and a positive second derivative with respect to the SNR. If all the successive derivatives of the SER also alternate in sign (referred to as complete monotonicity), then it is possible to express the SER as a positive mixture of decaying exponentials, which has applications in SER analysis over fading channels, as described in \cite{paper:STpaper11}. This serves as the motivation to explore the complete monotonicity ({\cm}) properties of the SER of arbitrary multi-dimensional constellations. An overview of the literature addressing related properties of the SER follows next. 

One-dimensional and two-dimensional constellations have been adopted in many communication systems in the literature, and investigations into the properties of the SER of these constellations have revealed the convexity of the SER with respect to the signal-to-noise ratio (SNR) under impairing additive white Gaussian noise (AWGN) \cite{paper:azizoglu96,paper:loyka10}. Some special cases of two dimensional constellations such as $M$-ary phase shift keying ($M$-PSK) and $M$-ary quadrature amplitude modulation ($M$-QAM) have SERs which are known to be completely monotone functions of the SNR \cite{paper:STpaper11, nesenbergs67}, which is a stronger condition than convexity. On the other hand, constellations of dimensionality greater than two (which we refer to as ``higher dimensional constellations'' henceforth) have found practical applications in satellite communications \cite{biglieri92, Taricco93} and more recently, in optical communications \cite{bulow09, agrell09}. Investigations of the convexity properties of the SER of such constellations are relatively scarce in the literature. It is known that the second derivative of the SER of a constellation of dimensionality greater than two is non-negative at sufficiently high SNR \cite{paper:loyka10}. Although this result is a general one, it does not provide a conclusive means of determining whether a given arbitrary constellation has a convex SER or not. For certain higher dimensional constellations, analytical expressions for the SER have been derived in the literature, which can be used to deduce the convexity of the SER. For example, the class of constellations of dimensionality $2,3,4$ and $5$ described in \cite{paper:agrell11} can be verified to have convex SERs under AWGN, by differentiation of the analytical SER expressions given in \cite{paper:agrell11}. On the other hand, verifying if a SER is completely monotone is difficult, even if a closed form expression for the SER is available. It has been established recently, that if the rank of a constellation matrix is at most two, then it will have a completely monotone SER under AWGN with ML detection \cite{paper:AdithyaISIT12}. However, there has been no investigation into the complete monotonicity properties of the SER for higher dimensional constellations, which can be used to simplify the expressions for average SERs over fading channels, and to establish useful comparisons of average SERs of a system under two different fading channels \cite{paper:STpaper11}, using the tools of stochastic ordering, which is surveyed next.

The theory of stochastic orders provides a comprehensive framework to
compare two random variables (RVs) \cite{book:shaked94}, and can be applied to compare two fading distributions based on a range of performance metrics such as the ergodic capacity or the SER. One example is the convex order, which finds applications in comparing the variability of two RVs by comparing the expectation over all convex functions over the two different distributions that are being compared \cite{paper:STpaper11}. There are many other stochastic orders that capture comparisons of RVs in terms of size and variability. These include the Laplace transform (LT) order \cite{book:shaked94} which compares the real valued Laplace transform of the probability density function (PDF) of two positive RVs. The instantaneous SNRs corresponding to many fading envelope distributions such as Nakagami and Ricean exhibit LT ordering with respect to their respective line of sight parameters \cite{paper:STpaper11}. This ordering can be systematically exploited through its connection with {\cm} functions, yielding generic comparisons of averages of a {\cm} function (such as SER) over two different positive random variables, even in cases where closed-form expressions for these averages are unavailable.

In this work, it is shown that the SER of an arbitrary multi-dimensional constellation impaired by additive independent and identically distributed ({\iid}) Gaussian noise under maximum likelihood (ML) detection can be represented as a product of a completely monotone function of SNR, and a power of SNR, which depends only on the rank of the constellation matrix. This result also generalizes to SERs under compound Gaussian noise, which includes many non-Gaussian noise distributions such as Middleton class-A noise \cite{paper:middleton97} and symmetric alpha-stable noise \cite{book:taqqu}. The SER of an arbitrary multi-dimensional constellation is shown to be completely monotone if the constellation matrix has a rank of one or two. Since complete monotonicity implies convexity, the SER is a convex function of the SNR, provided that the constellation matrix has a rank of one or two. For constellations matrices whose rank is greater than two, it is shown that the complete monotonicity of the SER depends on the constellation geometry and choice of prior probabilities. This work also describes a novel stochastic order for fading distributions, which can be used to order the average SERs of arbitrary multidimensional complex constellations over quasi-static fading channels, and generalizes the existing Laplace transform order on random variables.

The rest of the paper is organized as follows: Section \ref{sec:Preliminaries} surveys the relevant mathematical background, including a summary of stochastic ordering theory. Section \ref{sec:sys_model} describes the system model. The result on the representation of the SER of an arbitrary multi-dimensional constellation is detailed in Section \ref{sec:err_rate_multidim}. In Section \ref{sec:applications}, the applications such as (i) ordering the average SERs of constellations with {\cm} SERs over two different fading channels, and (ii) comparing the average SER of an arbitrary constellation using a newly proposed stochastic order, are discussed. Conclusions are presented in Section \ref{sec:conclusions}. 

Here are some remarks on the notations used in this paper. Vectors are denoted by boldface lower-case letters. Sets are denoted using upper-case script font. The set of real and natural numbers are denoted by $\set{R}$ and $\set{N}$ respectively. Transpose of a vector is denoted by $(\cdot)^{\rm T}$. The multivariate real (circularly symmetric complex) Gaussian distribution with mean vector $\vect{a}$, and covariance matrix $\mat{C}$ is denoted by $\mathcal{N}(\vect{a},\mat{C})$ ($\mathcal{CN}(\vect{a},\mat{C})$). $\E{}{g(X)}$ is used to denote the expectation of the function $g(\cdot)$ over the distribution of the random variable $X$. The identity matrix is denoted by $\mathbf{I}$. $\Pei$ denotes the SER conditioned on the $i^{th}$ symbol of the constellation being transmitted. $\AvgPe$ denotes the SER averaged over the constellation. $\overline{P}_{\rm e}(\rho)$ denotes the average SER over a fading channel (where the averaging is over the constellation points and the fading channel statistics). $\vectornorm{\cdot}$ denotes the vector-$2$ norm for both real and complex vector spaces. The indicator function is defined as $I[x \in \set{S}] = 1$, if $x \in \set{S}$ and $0$, otherwise.
\section{Mathematical Preliminaries}
\label{sec:Preliminaries}
\subsection{Complete Monotonicity}
\label{subsec:math_cm}
A function $g : (0,\infty) \rightarrow \re$ is completely monotone (c.m.), if and only if it has derivatives of all orders which satisfy
\begin{align}
\label{eqn:cm_def}
(-1)^{n} \frac{\D ^{n}}{\D x^{n}} g(x) \geq 0 ,
\end{align}
for all $n \in \N \cup \lbrace 0 \rbrace$, where the derivative of order $n=0$ is defined as $g(x)$ itself. The celebrated Bernstein's theorem \cite{book:schilling10} asserts that, $g : (0,\infty) \rightarrow \re$ is c.m. if and only if it can be written as a mixture of decaying exponentials: 
\begin{equation}
\label{eqn:bernstein}
g(x) = \int\limits_{0}^{\infty} \exp(-u x)  \mu(u) \D u,
\end{equation}
where $\mu : [0,\infty) \rightarrow [0,\infty)$ is called as the \emph{representing function} of $g$, in this paper. It is straightforward to verify that c.m. functions are positive, decreasing and convex, and that positive linear combinations of c.m. functions are also c.m.

A function $g: (0,\infty) \rightarrow \re$ is said to be completely monotone of order $\alpha \in \N$ if and only if $x^{\alpha} g(x)$ is {\cm}. If $g$ is {\cm} of order $\alpha$, then $g$ is also {\cm} of order $\beta$, where $0 \leq \beta < \alpha$. In \cite[Theorem 1.3]{paper:koumandos09}, it is shown that a necessary and sufficient condition for $g$ to be {\cm} of order $\alpha$ is that $g(x)$ must be represented in the form \eqref{eqn:bernstein}, where the integral converges for all $x>0$. In addition, $\mu$ must be $\alpha-1$ times differentiable, with the $k^{th}$ derivative of $\mu(u)$ equal to zero at $u=0$ for $0 \leq  k \leq \alpha-2$, and $\D^{\alpha-1} \mu(u)/\D u^{\alpha-1}$ nonnegative, right-continuous and non decreasing. 
\subsection{Stochastic Orders}
\label{sec:stoch_order_prelim}
Let $\F$ denote a set of real valued functions $g : \re^{+}  \rightarrow \re$, and $X$ and $Y$ be non-negative RVs. The integral stochastic order with respect to a set of functions $\F$ is defined as \cite{mullerbook_02}:
\begin{align}
\label{eqn:integral_st_order_def}   
X \leq_{\F} Y \Leftrightarrow \E{}{g(X)} \leq \E{}{g(Y)} \;,\; \forall g \in \F\; ,
\end{align}
where $\F$ is known as a generator of the order $\orderl{\F}$. We write $X \norderl{\F} Y$, if \eqref{eqn:integral_st_order_def} does not hold. Two examples of integral stochastic orders relevant to this paper are now considered, by specifying the corresponding generator set of functions $\F$.
\subsubsection{Convex Order}
In this case, $\F$ is the set of all convex functions, and the order is denoted as $X \orderl{cx} Y$. Applying \eqref{eqn:integral_st_order_def} using $g(x)=x$ and $g(x)=-x$ (both of which are convex), we have $\E{}{X} = \E{}{Y}$ whenever $X$ and $Y$ are convex ordered. Therefore, convex ordering establishes that the RVs have the same mean, and $X$ is ``less variable'' than $Y$. 

Instantaneous SERs of two-dimensional modulations over fading channels are known to be convex functions of the instantaneous SNR \cite{paper:loyka10}. Therefore, establishing convex ordering of two RVs can enable comparisons of the average performance of the same system subject to fading modeled by two different RVs.

\subsubsection{Laplace Transform Order}

This partial order compares random variables based on their Laplace transforms. In this case, $\F = \lbrace g(x) \mid g(x) = - \exp\left( - \rho x \right) , \; \rho \geq 0 \rbrace$, so that $X_{1} \orderl{Lt} X_{2}$ if and only if
\begin{align}
\label{eqn:LT_exp_order}
\E{}{\exp(-\rho X_{2})} \leq \E{}{\exp(-\rho X_{1})}, \; \forall \; \rho > 0 \;.
\end{align}
One useful property of LT ordered random variables, which can be obtained using \eqref{eqn:LT_exp_order} and \eqref{eqn:bernstein} is that $X_{1} \orderl{Lt} X_{2}$ is equivalent to
\begin{align}
\label{eqn:LT_cm_order}
\E{}{g(X_{2})} \leq \E{}{g(X_{1})},
\end{align}
for all {\cm} functions $g(\cdot)$. In other words, the generator $\F$ can be replaced by the set containing the negative of all {\cm} functions without changing the stochastic order \cite{mullerbook_02}. In a wireless communications context, let $\rho >0$ represent the average SNR, and $\rho X_{1}$ and $\rho X_{2}$ represent the instantaneous end-to-end SNRs of a system over two different fading channels with $X_{1} \orderl{Lt} X_{2}$, such as the case when $\sqrt{X_{1}}$ and $\sqrt{X_{2}}$ are Nakagami distributed with parameters $m_{1}$ and $m_{2}$ respectively, where $m_{1} \leq m_{2}$ \cite{paper:STpaper11}. Let $P_{\rm e}(\rho x)$ be the instantaneous SER of a modulation scheme, and let $P_{\rm e}(\rho x)$ be {\cm}. For example, $P_{\rm e}(\rho x) = (1/2)\exp(-\rho x)$ for the case of differential PSK modulation, or $P_{\rm e}(\rho x) = \Q{}{\sqrt{2 \rho x}}$ for BPSK (where $\Q{}{\sqrt{x}} := \int_{\sqrt{x}}^{\infty} (1/\sqrt{2 \pi}) \exp(-t^{2} /2) \D t$). Then using \eqref{eqn:LT_cm_order}, it is seen that $\E{}{P_{\rm e}(\rho X_{2})} \leq \E{}{P_{\rm e}(\rho X_{1})}$, $\forall \rho > 0$. This is because both choices of $P_{\rm e}(\rho x)$ are {\cm} functions for all $\rho$, and \eqref{eqn:LT_cm_order} is used with $g(x) = P_{\rm e}(\rho x)$. In other words, \eqref{eqn:LT_cm_order} can be used to compare the average SERs of the two channels at all SNR.

\subsection{Polyhedra and Polytopes}
A set $\set{P} \subseteq \re^{N}$ is a polyhedron if it is the intersection of finitely many closed half-spaces, i.e., $\set{P} := \lbrace \vect{x} | \mathbf{A}\vect{x} \leq \vect{b} \rbrace$, for some $ \mathbf{A} \in \re^{M \times N}$, $\vect{b} \in \re^{N}$, and the inequality is applied component-wise \cite{book:ziegler95}. A bounded polyhedron is referred to as a polytope. In a digital communications context, it is known that the decision region of a multi-dimensional constellation impaired by white Gaussian noise is a polyhedron.

A face of a polyhedron $\set{P}$ is the intersection of $\set{P}$ with a supporting hyperplane of $\set{P}$.
The dimension of a face is defined as the dimension of its affine hull. Faces of $\set{P}$ of dimension zero, dimension one, and dimension $N-1$ are known as vertices, edges and facets of $\set{P}$, respectively. 

Some examples of polyhedra relevant to this paper are described next. A polyhedral cone is a polyhedron, which is defined as ${\rm cone}(\set{Y}) := \lbrace \sum_{i=1}^{M} \lambda_{i} \vect{y}_{i} \mid \vect{y}_{i} \in \set{Y}, \lambda_{i} \geq 0 \rbrace$, where $\set{Y}$ is a non-empty set of points in $\re^{N}$. If the elements of $\set{Y}$ are linearly independent, then the polyhedral cone is called a simplicial cone. A well known result in combinatorial geometry literature is that any polyhedral cone admits a decomposition into simplicial cones \cite[Lemma 1.40]{book:de2010}. In this context, a decomposition of a polyhedron $\set{P}$ is defined as a collection of sets $\lbrace \set{P}_{1},\ldots,\set{P}_{R} \rbrace$, such that $\cup_{r=1}^{R} \set{P}_{r} = \set{P}$, and the intersection of any two sets in the decomposition is a common face of both, or the empty set.

\section{System Model}
\label{sec:sys_model}
In this paper, the transmission of an $N$-dimensional baseband signal $\rvect{s}$ through AWGN is considered, which is described as follows:
\begin{align}
\label{eqn:system_model} 
\rvect{y} = \rvect{s} + \rvect{z} \; ,
\end{align}
where the transmitted symbol $\rvect{s}  \in  \re^{N} $ is chosen from $\set{S} := \lbrace \vect{s}_{1},\ldots,\vect{s}_{M} \rbrace$. The \emph{constellation matrix} corresponding to $\set{S}$ is defined as $\mat{S} := [\vect{s}_{1},\ldots,\vect{s}_{M}]$, and the \emph{reduced dimension} of $\mat{S}$ is defined as the rank of $\mat{S}$, which is denoted by $\redN$. In \eqref{eqn:system_model}, the noise is assumed to be $\rvect{z} \sim \mathcal{N}(\vect{0},(1/\rho) \mathbf{I})$, $\rho >0$. When the signal energy is normalized as $M^{-1} \sum_{i=1}^{M} \vectornorm{\vect{s}_{i}} ^{2} =1$, then $M^{-1} \sum_{i=1}^{M} \vectornorm{\vect{s}_{i}} ^{2} /\E{}{\vect{z}^{\rm T}\vect{z}} = \rho$ represents the average SNR per dimension. At the receiver, the ML detector under AWGN is assumed, where the detected symbol $\hat{\vect{s}}$ is given by:
\begin{align}
\label{eqn:mdr_def}
\hat{\vect{s}} = \underset{\vect{s} \in \set{S}}{{\rm argmin}} \vectornorm{ \vect{y} - \vect{s} }^{2} \; ,
\end{align}
which is the ML detector for white Gaussian noise. Assuming that the origin of the coordinate system is shifted to $\vect{s}_{i}$, the Voronoi region of $\vect{s}_{i}$, denoted by $\Omegai$ is given by 
\begin{align}
\label{eqn:voronoiRegion}
\Omegai := \lbrace \vect{x} \in \re^{N} | \mat{A}_{i}\vect{x} \leq \vect{b}_{i} \rbrace ,
\end{align}
where $\mat{A}_{i} \in \re^{F \times N}$, where $F \leq M$, and the $j^{th}$ row of $\mat{A}_{i}$ is $\vect{a}_{j,i}^{\rm T} = (\vect{s}_{j} - \vect{s}_{i})^{\rm T}/\vectornorm{\vect{s}_{j} - \vect{s}_{i}}$, while the $j^{th}$ element of $\vect{b}_{i}$ is $b_{j,i} = \vectornorm{ \vect{s}_{j} - \vect{s}_{i} } /2 $. It is assumed that \eqref{eqn:voronoiRegion} is a non-redundant description of $\Omegai$. The minimum distance $d_{\rm min}$ of the constellation is defined through its square as $d_{\rm min}^{2} := \underset{\vect{s}_{i}, \vect{s}_{j} \in \set{S} ; \vect{s}_{j} \neq \vect{s}_{i}}{\min} \vectornorm{\vect{s}_{i} - \vect{s}_{j}}^{2}=4 \underset{i,j}{\min}\;b_{j,i}^{2}$ \cite{book:boyd04}.

The probability of error $\Pei$, given that $\vect{s}_{i}$ is transmitted is given by
\begin{align}
\label{eqn:pei_generic}
\Pei := \left(\frac{\rho}{2 \pi} \right)^{N/2} \int\limits_{\re^{N} - \Omegai} \exp\left[-\frac{\rho}{2} \vect{x}^{{\rm T}} \vect{x} \right] \D \vect{x} \;,
\end{align}
where $\set{S}_{1} - \set{S}_{2} := \lbrace \vect{x} \in \set{S}_{1} | \vect{x} \notin \set{S}_{2} \rbrace$ is the relative complement of $\set{S}_{2}$ in $\set{S}_{1}$. The probability of error averaged across all possible transmitted symbols is given by
\begin{align}
\label{eqn:A1_Pei_Pe}
\AvgPe = \sum_{i=1}^{M} \Pr[\vect{s} = \vect{s}_{i}] \Pei \;,
\end{align}
where $\Pr[\vect{s} = \vect{s}_{i}]$ represents the \emph{a priori} probability of transmitting $\vect{s}_{i}$. 

For any constellation $\mat{S}$, there exists an equivalent full rank constellation $\mat{S}^{*}$, which has the same SER as that of $\mat{S}$. Such a definition is useful in developing a representation for the SER of a multidimensional constellation. 
\begin{defn}[Reduced Constellation]
\label{def:red_const}
Let $\mat{S} = \mat{U} \mat{\Sigma} \mat{V}^{\rm T}$ be a singular value decomposition of $\mat{S}$, where $\mat{U} \in \re^{N \times N}$, $\mat{\Sigma} \in \re^{N \times M}$ and $\mat{V} \in \re^{M \times M}$, and the diagonal matrix consisting of the first $\redN$ rows and first $\redN$ columns of $\mat{\Sigma}$ contains the non-zero singular values of $\mat{S}$. Then the $\redN \times M$ matrix $\mat{S}^{*}$ given by the first $\redN$ rows of $\mat{\Sigma} \mat{V}^{\rm T}$ is defined as the reduced constellation of $\mat{S}$. 
\end{defn}
By definition, the number of rows of $\mat{S}^{*}$ is no greater than that of $\mat{S}$. In addition, $\mat{S}^{*}$ can be shown to have the same SER as $\mat{S}$. To see this, recall that the SER of the minimum distance detector depends only on the distance between the column vectors of $\mat{S}$ and the Frobenious norm of $\mat{S}$. It then suffices to show that the columns of $\mat{S}^{*}$ have the same distance properties and norm as that of $\mat{S}$. To this end, observe that $\mat{\Sigma} \mat{V}^{\rm T} = \mat{U}^{\rm T} \mat{S}$ has the same distance and norm properties as that of $\mat{S}$, since it is an orthogonal transformation on $\mat{S}$. Further, by construction, the last  $N- \redN$ rows of $\mat{\Sigma} \mat{V}^{\rm T}$ are zeros. Hence, the distance and norm properties of $\mat{\Sigma} \mat{V}^{\rm T}$ and $\mat{S}^{*}$ are identical. In addition, since $\vect{z}$ is AWGN, multiplying $\vect{z}$ by an orthogonal matrix does not change its statistics. It thus follows that the SER of $\mat{S}^{*}$ and $\mat{S}$ are identical. To conclude the discussion of reduced constellations, an example of a constellation and its reduced constellation is provided. Consider $\set{S} = \lbrace \vect{s}_{1}, \vect{s}_{2} \rbrace$, where $\vect{s}_{1} = [\sqrt{0.5} \; \sqrt{0.5}]^{\rm T}$, and $\vect{s}_{2} = [-\sqrt{0.5} \; -\sqrt{0.5}]^{\rm T}$. Using the definition of the reduced constellation, it is straightforward to see that the reduced constellation corresponding to $\set{S}$ is the BPSK constellation set, and therefore the SER of $\set{S}$ is identical to that of BPSK.
\section{Symbol Error Rates of Multi-Dimensional Constellations}
\label{sec:err_rate_multidim}

Throughout the paper, the focus is to obtain a functional characterization of the SER of a multidimensional constellation, rather than to obtain bounds or closed-form expressions for the SER. Such a characterization can be used to uncover its convexity and complete monotonicity properties. 

\subsection{Symbol Error Rates of Real Constellations Under AWGN}
To begin with, it is assumed the the transmitted symbol is a real vector, and the additive noise is white Gaussian. For constellations with reduced dimension $\redN =1$, the SER of the detector \eqref{eqn:mdr_def} under AWGN can be seen to be a positive linear combination of {\cm} functions of the form $\Q{}{\sqrt{2 \rho \eta}}, \eta >0$, which is {\cm}. The functional structure of the SER of constellations with $\redN \geq 2$ is addressed in Theorem \ref{thm:n_dim_error_rate}, whose proof requires a result from the combinatorial geometry literature, which is stated next. 

\begin{lem}
\label{lem:polytope_partition}
Let $\set{P}$ be an $N$-dimensional polyhedron in $\re^{N}$ with $F$ facets. If $\set{P}$ contains the origin in its interior, then $\re^{N}$ admits the decomposition $\set{X} :=\lbrace \set{D}_{f,q_{_{f}}} \rbrace_{f,q_{_{f}}}$, $f=1,\ldots,F$, $q_{_{f}} = 1,\ldots,Q_{_{f}}$, where $\set{D}_{f,q_{_{f}}}$ are $N$-dimensional simplicial cones, and $f$ can be viewed as an index of the facets of $\set{P}$.
\end{lem}
\begin{IEEEproof}
An outline of the proof is provided in Appendix \ref{app:lemma_proof_abridged}.
\end{IEEEproof}
In other words, using the facets of an $N$-dimensional polyhedron $\set{P} \subseteq \re^{N}$ which contains the origin, it is possible to decompose $\re^{N}$ into a collection of $N$-dimensional simplicial cones. In what follows, the representation theorem is stated.
\begin{theorem}
\label{thm:n_dim_error_rate} 
For a constellation $\set{S} \subseteq \re^{N}$, whose reduced constellation is $\set{S}^{*}$ and reduced dimension is $\redN \geq 2$, the SER of the detector \eqref{eqn:mdr_def} under AWGN admits the representation
\begin{align}
\label{eqn:th_main}
\AvgPe = \rho^{p} f_{{\rm cm}}(\rho)\;,
\end{align}
where $\fcm$ is {\cm}, and $p \geq \redN /2 -1$. In \eqref{eqn:th_main}, the representing function of $\fcm$ satisfies $\mu(u) =0$ when $u < d_{\rm min}^{2}/4$, and $\mu(u) \geq 0$ otherwise, where $d_{\rm min}$ is the minimum distance of the constellation.
\end{theorem}
\begin{IEEEproof}
See Appendix \ref{appendix:AWGN_main_th}.  
\end{IEEEproof}
To prove Theorem \ref{thm:n_dim_error_rate}, we work with a reduced constellation $\mat{S}^{*}$ that is full rank. This is needed since Lemma \ref{lem:polytope_partition} is used in Theorem \ref{thm:n_dim_error_rate}, where $\set{P}$ is a Voronoi region with dimension $\redN$. This highlights the need to work with the reduced constellation and dimension. The proof of Theorem \ref{thm:n_dim_error_rate} provided in Appendix \ref{appendix:AWGN_main_th} can be viewed as a generalization of the method adopted in \cite{paper:craig91} to obtain an expression for the SER of arbitrary two-dimensional constellations under AWGN. 

As an example to corroborate Theorem \ref{thm:n_dim_error_rate}, consider the square $M$-QAM constellation under AWGN. For this constellation, it is known that $d_{\rm min} = \sqrt{2}$ and $\AvgPe = \omega_{1} \Q{}{\sqrt{\eta \rho}}-\omega_{2}\Q{2}{\sqrt{\eta \rho}}$, with $\omega_{1}= 4 (\sqrt{M}-1)/ \sqrt{M}, \eta = 3/(M-1)$ and $\omega_{2} = \omega_{1}^{2}/4$ \cite{simon_alouini05}. In this case, $\AvgPe$ can be represented in the form \eqref{eqn:th_main}, where $p=0$ (since $\redN = 2$), and the representing function of $f_{\rm cm}(\rho)$ is given by
\begin{align}
\mu(u) = \frac{\sqrt{\eta}}{2 \pi}\left[ \frac{\omega_{1}I[0.5 \leq u \leq 1]}{u \sqrt{2 u -1}} + \frac{(\omega_{1}-\omega_{2})I[u \geq 1]}{u \sqrt{2u-1}} \right] \;,
\end{align}
which is zero when $u < 0.5$ and non-negative when $u \geq 0.5$, because $\omega_{1}-\omega_{2} > 0$. Further, since $p=0$, it is obvious that the square $M$-QAM constellation is an example of a constellation which has a {\cm} SER. 

According to Theorem \ref{thm:n_dim_error_rate}, the SER of every constellation can be written as $\AvgPe = \rho^{p} f_{\rm cm}(\rho)$, where $p = \redN/2 -1$ and $f_{\rm cm}$ is {\cm}. However, it is  possible that the SER of some constellations admit a representation of the form \eqref{eqn:th_main}, where the exponent of $\rho$ is less than $\redN/2 -1$. As a result, constellations for which this exponent is zero have {\cm} SERs. The following corollary of Theorem \ref{thm:n_dim_error_rate} establishes a necessary and sufficient condition for the SER of a constellation to be {\cm}.

\begin{corollary}
\label{thm:cm_BER_AWGN}
The SER of $\set{S}$ using the detector \eqref{eqn:mdr_def} under AWGN is {\cm} if its reduced dimension satisfies $\redN \leq 2$. Conversely, let $\mu(u) = \int_{0}^{\infty} \exp(\rho u) \rho^{-(\redN/2 -1)}  \AvgPe \D \rho$, $\hat{\mu}(u) = \int_{0}^{u} \mu(u-v) v^{-1/2} \D v$ and $r= \lceil \redN/2 -2 \rceil$. If $\redN > 2$ and $\redN$ is even (odd), the SER is {\cm} if and only if $\mu(u)$ $(\hat{\mu}(u))$ is $r$ times differentiable, and $\D^{k} \mu(u)/\D u^{k} \geq 0$ ($\D^{k} \hat{\mu}(u)/\D u^{k} \geq 0$) for $0 \leq k \leq r$, and  $\D^{r} \mu(u)/\D u^{r} (\D^{r} \hat{\mu}(u)/\D u^{r})$ is increasing and continuous.
\end{corollary}
\begin{IEEEproof}
See Appendix \ref{appendix:corr_1}.
\end{IEEEproof}
In other words, Corollary \ref{thm:cm_BER_AWGN} states that the complete monotonicity of the SER for constellations with $\redN \leq 2$ does not depend on the geometry of the constellation. However, for constellations with higher reduced dimensions, the complete monotonicity of the SER depends on the differentiability of the representing function corresponding to $f_{\rm cm}(\rho)$, which is a function of the constellation geometry and the a-priori probabilities. Although Corollary \ref{thm:cm_BER_AWGN} is applicable to any constellation, it is not easy to obtain the equivalent set of conditions on the constellation geometry and prior probabilities under which the SER of a constellation with $\redN > 2$ is {\cm}, and this is posed as an open problem. 

Next, we provide instructive examples through which the complete monotonicity of the SER can be seen to depend on the constellation geometry for $\redN = 3$. First, consider the constellation where the points are chosen as the vertices of a cube. In this case, the SER is given by $\AvgPe = 1-(1-\Q{}{\sqrt{2 \rho}})^{3}$, which can be rewritten in the form \eqref{eqn:bernstein}, with $\mu(u)= (3/\pi)I[1 \leq u \leq 2]+(\pi-\cos^{-1}(\alpha(u)))/2\pi^{2} I[3 \leq u \leq 4]+ (\pi + \cos^{-1}(\alpha(u)))/2\pi^{2} I[u \geq 4]$, where $\alpha(u) = (3 u^{2} - 12u +8)/(u-2)^{3}$. Since $\mu(u) \geq 0$, from Bernstein's theorem, $\AvgPe$ is {\cm}. On the other hand, consider the $3$-D square QAM constellation, whose points are given by all possible sign permutations of $(\pm 1/\sqrt{6},\pm 1/\sqrt{6},\pm 1/\sqrt{6})$ and $(\pm 1/\sqrt{2}, \pm 1/\sqrt{2}, \pm 1/\sqrt{2})$. For this case, under the assumption of equal prior probabilities, numerical evaluation of the SER shows non-convexity (Fig. \ref{fig:QAM3D}). As a result, from \eqref{eqn:cm_def}, the SER is not {\cm}. Therefore, the {\cm} properties of the SER of a constellation with $\redN >2$ depends on the geometry and prior probabilities. 

With reference to existing literature, Corollary \ref{thm:cm_BER_AWGN} is a useful generalization of \cite[Theorem 2]{paper:loyka10}, which does not address complete monotonicity, or the possibility of reduced dimension. Further, from Corollary \ref{thm:cm_BER_AWGN}, it follows that the SER of any two-dimensional constellation under AWGN is convex. This particular consequence of Corollary \ref{thm:cm_BER_AWGN} has been previously established in \cite[Theorem 1]{paper:loyka10} using a different approach. In what follows, the behavior of second derivative of the SER for constellations with $\redN > 2$ is studied.

\begin{corollary}
\label{thm:high_dim_cvx}
If the reduced dimension $\redN$ of a constellation $\set{S}$ is greater than two, then the SER of the detector \eqref{eqn:mdr_def} under AWGN satisfies $\AvgPedd \geq 0$ when $\rho \geq \rho_{0}$, where $\rho_{0} :=4(p+\sqrt{p})/d_{\rm min}^{2}$, $p = \redN /2 -1$, and $d_{\rm min}$ is the minimum distance of the constellation.
\end{corollary}
\begin{IEEEproof}
See Appendix \ref{appendix:high_dim_cvx}
\end{IEEEproof}

A similar but weaker result has been obtained in \cite{paper:loyka10}, where $\rho_{0} = 4(N + \sqrt{N})/d_{\rm min}^{2}$. Corollary \ref{thm:high_dim_cvx} provides a larger region where the SER has positive second derivative, which is an improvement over \cite{paper:loyka10}, since $ \redN/2 -1  < \redN \leq N$. Although Corollary \ref{thm:high_dim_cvx} establishes a bound on the values of SNR for which the second derivative becomes non-negative, it does not forbid $\AvgPedd \geq 0$ for $\rho < \rho_{0}$. Indeed, it is possible for some multi-dimensional constellation with $\redN > 2$ to posses a convex SER. An example of such a constellation is one with $N=3$, wherein the points correspond to the vertices of regular convex polytope (RCP) \cite{paper:agrell11}, for which convexity follows from the expression for the SER given by \cite[Eqn. (2) - (7)]{paper:agrell11}. 

\subsection{Extension to Compound Gaussian Noise}
\label{sec:otherNoise}
In what follows, Theorem \ref{thm:n_dim_error_rate} and its corollaries are generalized to the case of additive compound Gaussian noise. 

The system model considered is still as in \eqref{eqn:system_model}, except that the additive noise is assumed to be $\rvect{Z} = \sqrt{W}\rvect{G}$, where $W$ is a positive RV, which is independent of each component of $\rvect{G} := [G_{1},\ldots,G_{N}]^{\rm T}$, and $G_{k} \sim \mathcal{N}(0,1/\rho)$ are iid, for $k=1,\ldots,N$. It should be noted that the elements of $\rvect{Z}$ are statistically dependent but uncorrelated in this case. Depending on the distribution of $W$, a number of noise distributions of interest arise from this formulation. For example, when $W$ is an affine function of a Poisson RV, $\vect{z}$ follows a Middleton class-A distribution \cite{paper:middleton97}, which is used to model multi-user interference; if $W$ is a positively skewed alpha-stable RV with a characteristic function $\varphi_{_{W}}(\omega) = \exp[-|\omega|^{\alpha}(1-j{\rm sgn}(\omega)\tan \pi \alpha/2)]$, where $j = \sqrt{-1}$, $0 < \alpha < 1$, and ${\rm sgn}(x)$ is the sign of $x$, then $\vect{z}$ follows the symmetric alpha-stable distribution \cite{book:taqqu}, with characteristic function $\varphi_{_{Z}}(\omega) = \exp[-|\omega|^{2 \alpha} ]$. For different values of $\alpha$, many impulsive noise distributions are obtained. For example, when $\alpha=1/2$, $\vect{z}$ is Cauchy distributed noise \cite{book:taqqu}. 

In this discussion, the minimum distance detector \eqref{eqn:mdr_def}, which is still the maximum likelihood detector for dependent but uncorrelated compound Gaussian noise, is assumed to be used at the receiver side. The reduced constellation corresponding to $\set{S}$ is defined as in Definition \ref{def:red_const} and is denoted by $\set{S}^{*}$. It is not difficult to show that the SER of $\set{S}$ and $\set{S}^{*}$ are identical under additive compound Gaussian noise.

An extension of Theorem \ref{thm:n_dim_error_rate} to the case of additive compound Gaussian noise is now developed. Conditioning on $W=w$, $\rvect{Z}$ is an i.i.d multivariate Gaussian, in which case Theorem \ref{thm:n_dim_error_rate} can be invoked to get $P_{\rm e}(\rho | W=w) = \rho^{p} f_{\rm cm}(\rho ; w)$, where $f_{\rm cm}(\rho ; w)$ is {\cm} in $\rho$ for each $w$. Averaging over the distribution of $W$, the equivalent of Theorem \ref{thm:n_dim_error_rate} for the case of compound Gaussian noise is obtained, since the expectation can be interpreted as a positive linear combination of $f_{\rm cm}(\rho ; w)$, which results in a {\cm} function, denoted by $f_{\rm cm}(\rho)$.

Extensions of the corollaries of Theorem \ref{thm:n_dim_error_rate} for the case of compound Gaussian noise are also seen to be true, since they are obtained from the generalization of Theorem \ref{thm:n_dim_error_rate} to this noise model, without any additional assumptions.

\subsection{Extension to Complex Constellations Under AWGN}
\label{subsec:complexExt}
Theorem \ref{thm:n_dim_error_rate} and its corollaries have been derived under the assumption that the transmitted symbol is chosen from a real constellation. A system model which is more relevant to in-phase/quadrature communication schemes and communication over fading channels is one where the transmitted symbol is a complex vector. Motivated by this, an extension of Theorem \ref{thm:n_dim_error_rate} and its corollaries are obtained for the system model \eqref{eqn:system_model}, where $\vect{s}$ is chosen from  $\set{S} = \lbrace \vect{s}_{1},\ldots,\vect{s}_{M} \rbrace$ with $\vect{s}_{i} \in \mathcal{C}^{N}$, $i = 1, \ldots, M$, and the additive noise $\vect{z} \sim \mathcal{CN}(\vect{0}, (2/\rho)\mathbf{I})$. The receiver assumed for this discussion is the ML detector under circularly symmetric complex Gaussian noise, which is given by \eqref{eqn:mdr_def}, where the $\vectornorm{\cdot}$ in \eqref{eqn:mdr_def} is interpreted as the $2$-norm of a complex vector. Before proceeding to develop an extension of Theorem \ref{thm:n_dim_error_rate} to the complex case, a useful result pertaining to the SER of complex constellations is noted.
\begin{theorem}
\label{lem:SER_complex_real}
Let $\Pei$ denote the SER of \eqref{eqn:system_model} conditioned on the transmission of $\vect{s}_{i} \in \set{S} := \lbrace \vect{s}_{1},\ldots,\vect{s}_{M} \rbrace$, where $\vect{s}_{j} \in \mathcal{C}^{N}$, $j= 1,\ldots,M$, and $\vect{z} \sim \mathcal{CN}(\vect{0}, (2/\rho)\mathbf{I})$. Further, let $\tilde{P}_{{\rm e},i}(\rho)$ denote the SER of \eqref{eqn:system_model} conditioned on the transmission of $\tilde{\vect{s}}_{i}  \in \tilde{\set{S}} := \lbrace \tilde{\vect{s}}_{1},\ldots,\tilde{\vect{s}}_{M} \rbrace$, where $\tilde{\vect{s}}_{j} := [{\rm Re} \lbrace \vect{s}_{j} \rbrace^{\rm T} \; {\rm Im}\lbrace \vect{s}_{j} \rbrace^{\rm T}]^{\rm T}$, $j=1,\ldots,M$, and $\vect{z} \sim \mathcal{N}(\vect{0}, (1/\rho)\mathbf{I})$. Then, $\Pei = \tilde{P}_{{\rm e},i}(\rho)$.
\end{theorem}
\begin{IEEEproof}
It is straightforward to show that the Voronoi regions corresponding to $\vect{s}_{i}$ and $\tilde{\vect{s}}_{i}$ are identical for $i = 1,\ldots, M$. The Theorem then follows by observing that the SER is the Gaussian integral outside the Voronoi region.
\end{IEEEproof}
Therefore, in order to generalize Theorem \ref{thm:n_dim_error_rate} to the case of a complex $N$-dimensional transmitted symbol vector $\vect{s} \in \set{S}$, it is sufficient to consider the transmission of a real $2N$-dimensional vector $\tilde{\vect{s}} \in \tilde{\set{S}}$, where the $i^{th}$ element of $\tilde{\set{S}}$ is given by $[{\rm Re} \lbrace \vect{s}_{i} \rbrace^{\rm T} \; {\rm Im}\lbrace \vect{s}_{i} \rbrace^{\rm T}]^{\rm T}$, with $\vect{s}_{i} \in \set{S}$ for $i = 1,\ldots, M$. Further, the SER $\tilde{\set{S}}$ is equal to that of its reduced constellation $\tilde{\set{S}}^{*}$ described in Definition \ref{def:red_const}, and the reduced dimension $\tilde{N}^{*}$ of $\tilde{\set{S}}$ is given by the rank of $\tilde{\mat{S}}$, the $i^{th}$ column of which is given by $[{\rm Re} \lbrace \vect{s}_{i} \rbrace^{\rm T} \; {\rm Im}\lbrace \vect{s}_{i} \rbrace^{\rm T}]^{\rm T}$, $i = 1,\ldots, M$. Since $\tilde{\set{S}}$ is a real constellation, the expression for the SER is given by Theorem \ref{thm:n_dim_error_rate}, where $\redN$ is replaced by $\tilde{N}^{*}$. From Theorem \ref{lem:SER_complex_real}, this is also the SER of a complex $N$-dimensional vector symbol set under complex AWGN. Using the same line of argument, it can be seen that the SER of a complex $N$-dimensional constellation is {\cm} in the SNR if and only if its reduced dimension $\tilde{N}^{*}$ is one or two, as in Corollary \ref{thm:cm_BER_AWGN}. Proceeding in a similar fashion, Corollary \ref{thm:high_dim_cvx} can also be generalized to the complex case, by replacing $\redN$ with $\tilde{N}^{*}$ in the corollary to obtain the same conclusions.

\section{Applications}
\label{sec:applications}
In this section, applications of Theorem \ref{thm:n_dim_error_rate} and Corollary \ref{thm:cm_BER_AWGN} in the context of ordering of wireless system performance is presented.

\subsection{Applications in Stochastic Ordering}
\label{sec:st_order_fading}
Complete monotonicity of SER for complex constellations with a reduced dimension of one or two, as suggested by extension of Corollary \ref{thm:cm_BER_AWGN} to complex constellations finds immediate application in comparing the average SER of such constellations over two different fading channels using the theory of stochastic ordering mentioned in Section \ref{sec:stoch_order_prelim}. To elucidate further, the following system model is considered:
\begin{align}
\label{eqn:fading_model}
\rvect{y} =  h\rvect{s}+\rvect{v} \;,
\end{align}
where the effect of quasi-static fading is captured by the complex scalar RV $h$ whose real and imaginary parts are independent of each other, $\rvect{s} \in \set{S}$ with $\vect{s} \in \mathcal{C}^{N}$, and $\rvect{z} \sim \mathcal{CN}(\vect{0},(1/\rho) \mathbf{I})$ is the circularly symmetric AWGN. For this system, the instantaneous channel gain is defined as $X:= | h |^{2}$, and the instantaneous SNR is given by $\rho X$. Assuming that the receiver has full channel state information, the instantaneous SER can be shown to be a function of the instantaneous SNR only, and the average SER is obtained by taking the expectation of the instantaneous SER over the distribution of the instantaneous channel gain. In this application, goal is to compare the average SER of the system \eqref{eqn:fading_model} under two different fading channels with instantaneous channel gains $X_{1}$ and $X_{2}$.

To begin with, let $\set{S}$ be a complex constellation with a reduced dimension less than or equal to two, which has {\cm} SER according to the complex extension of Corollary \ref{thm:cm_BER_AWGN}. Now, consider two fading scenarios with instantaneous channel gains $X_{1}$ and $X_{2}$, such that $X_{1} \orderl{Lt} X_{2}$. Then, according to \eqref{eqn:LT_cm_order}, $\E{}{P_{\rm e}(\rho X_{2})} \leq \E{}{P_{\rm e}(\rho X_{1})}, \forall \rho >0$. As a result, complete monotonicity of SER can be exploited to compare two fading channels at all SNR, based on the average SER, even in cases where the expression for the average SER is not analytically tractable. As an illustrative example, consider the quadrature-PSK (QPSK) constellation, for which the reduced dimension can be seen to be equal to $2$, and $\Pe{}{\rho x} = \Q{}{\sqrt{2 \rho x}}$ (assuming equal prior probabilities), which is {\cm} in $x$. Now, assume QPSK is used over two different Nakagami-$m$ fading channels, the first one with LoS parameter $m_{1}$ and instantaneous channel gain $X_{1}$, and the second one with LoS parameter $m_{2}$ and instantaneous channel gain $X_{2}$, where $m_{2} \geq m_{1}$ so that $X_{1} \orderl{Lt} X_{2}$. In this case, \eqref{eqn:LT_cm_order} implies that $\E{}{\Q{}{\sqrt{2 \rho X_{2}}}} \leq \E{}{\Q{}{\sqrt{2 \rho X_{1}}}}, \; \forall \rho>0$, which provides a way of comparing the average SERs over two fading channels with different LoS parameters.

The complex extension of Corollary \ref{thm:cm_BER_AWGN} suggests that the SER of a complex constellation with reduced dimension greater than or equal to four is not {\cm}, and thus the LT ordering of instantaneous channel gains of two fading channels does not provide a conclusive comparison of the average SER of these channels. Motivated by this, a new stochastic order is introduced next, which can be used to compare the average SER of a multidimensional complex constellation over two different complex fading channels. The stochastic order $\orderl{\F_{p}}$ is formally defined below.

\begin{defn}  
Let $X_{1}$ and $X_{2}$ be two positive RVs, and let $p \geq 0$ be fixed. Then $X_{1} \leq_{\F_{p}} X_{2}$ if and only if $\E{}{X_{1}^{p} \exp(-\rho X_{1})} \geq \E{}{X_{2}^{p} \exp(-\rho X_{2})}$ for all $\rho > 0$.
\end{defn}
In other words, for every $p \geq 0$, $\orderl{\F_{p}}$ is an integral stochastic order in the sense of \eqref{eqn:integral_st_order_def}, with $\F_{p} = \lbrace g(x) \mid g(x) = -x^{p} \exp(-\rho x) , \rho \geq 0\rbrace$. 
A necessary and sufficient condition for $X_{1} \leq_{\F_{p}} X_{2}$ can be proved to be as follows:
\begin{theorem}
\label{thm:high_ord_st_ord}
Let $X_{1}$ and $X_{2}$ be two positive RVs, and $p \geq 0$. Then, $X_{1} \leq_{\F_{p}} X_{2}$ if and only if
\begin{align}
\E{}{X_{1}^{p}f_{\rm cm}(X_{1})} \geq \E{}{X_{2}^{p}f_{\rm cm}(X_{2})} \;,
\end{align}
where $f_{\rm cm}(\cdot)$ is {\cm}.
\end{theorem}
\begin{IEEEproof}
See Appendix \ref{app:appendix_C}.
\end{IEEEproof}

Verifying the $\orderl{\F_{p}}$ order for any pair of random variables, when $p \in \set{N} \cup \lbrace 0 \rbrace$ is relatively straightforward, and can be done by comparing the $p^{th}$ derivative of the real-valued Laplace transforms of the densities of the two RVs. Clearly, $\leq_{\F_{p}}$ is the LT order, when $p=0$. In this case, the envelope fading distributions for $\sqrt{X_{1}}$ and $\sqrt{X_{2}}$ such as Nakagami-$m$ satisfy $X_{1} \leq_{\F_{0}} X_{2}$, when $m_{1} \leq m_{2}$ \cite{paper:STpaper11}. Intriguingly however, for any $p >0$, fading channels modelled using Nakagami-$m$ or Rician distributions do not satisfy the $\orderl{\F_{p}}$ order with respect to their corresponding line of sight parameters. For example, in the Nakagami-$m$ fading scenario, $X = |h|^{2}$ in \eqref{eqn:fading_model} is Gamma distributed. In this case, 
\begin{align}
\E{}{X^{p}\exp(-\rho X)} = \frac{m^m (m+ \rho)^{-m-p} \Gamma[m+p]}{\Gamma[m]}\;, 
\end{align}
which increases with $m$ for small $\rho$ and decreases otherwise, for any fixed $p>0$. Thus, if $p >0$, $X_{1} \norderl{\F_{p}} X_{2}$.  

Some implications of Theorem \ref{thm:high_ord_st_ord} to the ordering of average SERs of multidimensional constellations over fading channels are now considered. If $X_{1}$ and $X_{2}$ are the instantaneous channel gains of two fading scenarios characterized by \eqref{eqn:fading_model}, then according to Theorem \ref{thm:high_ord_st_ord}, it is easy to show that 
\begin{align}
\label{eqn:pe_order1}
X_{1} \leq_{\F_{p}} X_{2} \Rightarrow \E{}{\Pe{}{\rho X_{1}}} \geq \E{}{\Pe{}{\rho X_{2}}} \;, \forall \rho >0 \;,
\end{align}
where $\Pe{}{\cdot}$ is the instantaneous SER of a complex constellation $\set{S}$ with reduced dimension $\redN$, and $p =  \redN/2  -1 $. This is because, from the complex extension of Theorem \ref{thm:n_dim_error_rate} we have $\AvgPe = \rho^{p} f_{\rm cm}(\rho)$, which implies $\AvgPe \in \lbrace g(\rho) | g(\rho) = \rho^{p} f_{\rm cm}(\rho), p \geq 0 \;\rbrace$, and thus \eqref{eqn:pe_order1} follows from Theorem \ref{thm:high_ord_st_ord}. 

It has been reported in the literature that, it is possible to find a fading distribution such that the SER of the AWGN channel is worse than that under the fading case at low SNR, when higher dimensional constellations are employed \cite{paper:loyka10}. However, examples of such fading distributions have not been a subject of investigation. Using \eqref{eqn:pe_order1}, it is now shown that the Nakagami-$m$ fading case is an such an example. To begin with, consider the pure AWGN channel (i.e. the no fading scenario, where $\Pr[X_{1} = 1] = 1)$, for which $\E{}{X^{p}_{1} \exp(-\rho X_{1})} = \exp(-\rho)$. If $X_{2}$ denotes the instantaneous channel gain of a Nakagami-$m$ fading channel, then for every $0<m<\infty$, we now argue that there exists a $\rho_{1} >0$ such that $\E{}{P_{\rm e}(\rho X_{1})} \geq \E{}{P_{\rm e}(\rho X_{2})}$ for $\rho \leq \rho_{1}$, while $\E{}{P_{\rm e}(\rho X_{1})} \leq \E{}{P_{\rm e}(\rho X_{2})}$, for $\rho \geq \rho_{1}$. To this end, observe that $\E{}{X^{p}_{1} \exp(-\rho X_{1})}$ is greater than $\E{}{X^{p}_{2} \exp(-\rho X_{2})}$ when $\rho \leq \rho_{1}$, and vice-versa when $\rho \geq \rho_{1}$. Therefore, from \eqref{eqn:pe_order1}, the AWGN channel is worse than a Nakagami-$m$ channel in terms of SER of constellations with $\redN > 2$ at low SNR.

Next, a relation between $X_{1} \leq_{\F_{q}} X_{2}$ and $X_{1} \leq_{\F_{p}} X_{2}$ is obtained, where $p$ and $q$ are non-negative.
\begin{theorem}
\label{thm:order_vs_length}
Let $X_{1}$ and $X_{2}$ be two positive RVs. Then, for $0 \leq q \leq p$,
$
X_{1} \leq_{\F_{p}} X_{2} \Rightarrow X_{1} \leq_{\F_{q}} X_{2}.
$
\end{theorem}
\begin{IEEEproof}
Since $X_{1} \leq_{\F_{p}} X_{2}$, from Theorem \ref{thm:high_ord_st_ord}, we have $\E{}{X_{1}^{p}f_{\rm cm}(X_{1})} \geq \E{}{X_{2}^{p}f_{\rm cm}(X_{2})}$, for every {\cm} function $f_{\rm cm}(\cdot)$. Choose $f_{\rm cm}(x) := x^{-k} g_{\rm cm}(x)$, where $0 \leq k \leq p$, and $g_{\rm cm}(x)$ as some {\cm} function. Clearly, $f_{\rm cm}(x)$ as defined is {\cm}, since $x^{-k}$ is {\cm} for $k \geq 0$ and a product of {\cm} functions is also {\cm}. As a result, according to Theorem \ref{thm:high_ord_st_ord}, $X_{1} \leq_{\F_{p}} X_{2}$ implies $\E{}{X_{1}^{p-k}g_{\rm cm}(X_{1})} \geq \E{}{X_{2}^{p-k}g_{\rm cm}(X_{2})}$. The theorem then follows by assuming $q = p-k$. 
\end{IEEEproof}
Theorem \ref{thm:order_vs_length} in conjunction with \eqref{eqn:pe_order1} implies that if $X_{1} \leq_{\F_{p}} X_{2}$, then $X_{2}$ is better than $X_{1}$ in terms of average SERs of all constellations at all average SNR, with the reduced dimension of the constellation satisfying $ \redN /2 -1  \leq p$. It is interesting to investigate the conditions on $X_{1}$ and $X_{2}$ such that $X_{1} \leq_{\F_{p}} X_{2}$ for all $p \geq 0$. In that case, $X_{2}$ will be better than $X_{1}$ in terms of average SERs of any multi-dimensional constellation at all SNRs. However, this condition is not satisfied by any pair of random variables, as described in the following Theorem:
 \begin{theorem}
\label{thm:high_ord_st_ord_neg}
There are no two positive random variables which satisfy $X_{1} \leq_{\F_{p}} X_{2}$, for all $p \geq 0$.
\end{theorem}
\begin{IEEEproof}
See Appendix \ref{app:high_ord_st_ord_neg}.
\end{IEEEproof}
As described in the paragraph above Theorem \ref{thm:order_vs_length}, the AWGN channel is not the best channel in terms of SER of constellations with $\redN > 2$ at all SNR. According to Theorem \ref{thm:high_ord_st_ord_neg}, there is no fading distribution which dominates every other fading distribution in the sense of $\leq_{\F_{p}}$ for all $p \geq 0$. Therefore, unlike cases where the SER metric is convex, where AWGN (no fading) outperforms any fading, this is not the case for $\redN > 2$. Moreover, Theorem \ref{thm:high_ord_st_ord_neg} suggests that there is no fading distribution which serves the role of ``best'' fading distribution, in terms of SERs of constellations of every dimension. 
\section{Conclusions}
\label{sec:conclusions}
In this paper, the SER of an arbitrary constellation with reduced dimension $\redN \geq 2$ under AWGN is characterized as $\Pe{}{\rho} = \rho^{p}f_{\rm cm}(\rho)$, where $f_{\rm cm}(\cdot)$ is a {\cm} function, and $p \geq  \redN/2 -1 $. This representation of the SER is shown to apply to cases when the noise follows a compound Gaussian distribution. The expression for the SER obtained herein is useful in establishing that the SER is a {\cm} function if the constellation has a reduced dimension of one or two. The complete monotonicity of SER for constellations with $\redN >2$ is shown to depend on the differentiability  properties of the representing function corresponding to $\rho^{-p}\Pe{}{\rho}$, which is a function of the constellation geometry and the prior probabilities. The exact relation between the constellation geometry and the complete monotonicity of the SER for constellations with $\redN > 2$ is left as an open problem. Complete monotonicity of the SER has applications in obtaining comparisons of averages of SERs over pairs of quasi-static fading channels, such as Nakagami-$m$, whose instantaneous SNRs are Laplace transform ordered. Such comparisons can be made even in cases where a closed-form expression for the average SER is not analytically tractable. In addition, a new stochastic ordering relation is introduced, which can be exploited to obtain comparisons of the average SER of an arbitrary multidimensional constellation over two different fading channels.
\appendices

\section{Proof of Lemma \ref{lem:polytope_partition}}
\label{app:lemma_proof_abridged}
The proof of this lemma rests on the fact that it is possible to decompose $\re^{N}$ into a set $\set{F}$ consisting of polyhedral cones $\set{C}_{1},\ldots,\set{C}_{F}$, using the facets of an $N$-dimensional polyhedron $\set{P} \in \re^{N}$, which contains the origin in its interior \cite[pp. 192]{book:ziegler95}. Since every polyhedral cone admits a decomposition into $N$-dimensional simplicial cones \cite[Lemma 1.40]{book:de2010}, it is possible to decompose each $\set{C}_{f}$ into a set of $N$-dimensional simplicial cones $\lbrace \set{D}_{f,q_{_{f}}} \rbrace_{q_{_{f}}}$, for $f=1,\ldots,F$. Consequently, $\re^{N}$ admits a decomposition into $N$-dimensional simplicial cones, given by $\lbrace \set{D}_{f,q_{_{f}}} \rbrace_{f,q_{_{f}}}$.

\section{Proof of Theorem \ref{thm:n_dim_error_rate}}
\label{appendix:AWGN_main_th}
Throughout this appendix, we work with the reduced constellation $\mat{S}^{*}$, with rank $\redN$. Recall the AWGN system model \eqref{eqn:system_model}. To obtain an expression for the symbol error rate averaged over all constellation points $\AvgPe$, we first evaluate $\Pei$ given by \eqref{eqn:pei_generic}, and then use \eqref{eqn:A1_Pei_Pe}. For the sake of simplicity, we assume that the Voronoi region of $\vect{s}_{i}^{*} \in \mat{S}^{*}$ is a polytope. The following proof can easily be extended to cases when $\Omegai$ is an unbounded polyhedron, by assuming an additional facet $\vect{c}_{0} \vect{x} \leq 1$, which turns $\Omegai$ into a polytope \cite[pp. 75]{book:ziegler95}, and subsequently taking the limit of $\AvgPe$ so obtained as $\vect{c}_{0} \rightarrow 0$.  

We begin with an outline of the proof. In general, evaluating \eqref{eqn:pei_generic} is not straightforward, since $\Omegai$ is a polytope. However, the Gaussian integral in \eqref{eqn:pei_generic} can be simplified, if the region of integration is of the form $\set{Z} - \tilde{\set{Z}}$, where $\set{Z}$ is an $N$-dimensional simplicial cone, and $\tilde{\set{Z}}$ is the intersection of a halfspace and $\set{Z}$. To this end, we show that the Voronoi region of $\vect{s}_{i}^{*}$ has a dimension of $\redN$, so that Lemma \ref{lem:polytope_partition} can be used, and thus $\Pei$ in \eqref{eqn:pei_generic} can be rewritten as a sum of integrals over regions of the form $\set{Z} - \tilde{\set{Z}}$, where $\set{Z}$ and $\tilde{\set{Z}}$ are as defined above. Each of these integrals when expressed in hyperspherical coordinates yields a canonical structure, which can be algebraically manipulated to obtain \eqref{eqn:th_main} in the Theorem. In order to show that $\redN/2 -1$ in \eqref{eqn:th_main} is the smallest exponent for which the Theorem holds, an argument involving complete monotonicity of order $\alpha$ is provided towards the end of this appendix.  

In what follows, we present the details of the proof. Let $\Omegai$ be a non-redundant description of the Voronoi region of $\vect{s}_{i}^{*} \in \set{S}^{*}$, for $i=1,\ldots,M$. With a slight abuse of notation, we are dropping the superscript $*$ from $\Omegai$, to simplify the notation. First, we show that $\Omegai$ satisfies the conditions of Lemma \ref{lem:polytope_partition}. To this end, for any set of affinely independent points in space (such as $\set{S}^{*}$), the dimension of the Voronoi region corresponding to each point is equal to the dimension of the affine hull of the set ($\redN$, when the set of points is $\set{S}^{*}$) \cite[p. 232]{book:du95}. Therefore, $\Omegai$ is an $\redN$-dimensional polytope in $\re^{\redN}$. Also, since the origin of the coordinate system is shifted to $\vect{s}_{i}^{*}$, $\vect{0} \in \Omegai$. Thus, $\Omegaio$ satisfies the conditions of Lemma \ref{lem:polytope_partition}. As a result, using Lemma \ref{lem:polytope_partition} we obtain a set $\set{X}_{i}:= \lbrace \set{D}_{f,q,i} \rbrace_{q,f}$, which is a decomposition of $\re^{\redN}$ into $\redN$-dimensional simplicial cones (see Fig. 1). Clearly, every $\vect{x} \notin \Omegaio$ satisfies $\vect{x} \in \set{D}_{f,q,i} - \Omegaio$ for some $\lbrace q, f \rbrace$. Let the number of facets of $\Omegai$ be $F_{i}$. It now follows that
\begin{align}
\label{eqn:A1_Pe1_1_1}
\Pei = \dsum \J \;,
\end{align}
where 
\begin{align}
\label{eqn:A1_Pe1_2}
\J = \left( \frac{\rho}{2 \pi} \right)^{\redN/2} \int\limits_{\set{D}_{f,q,i} - \Omegaio}  \exp \left( -\frac{\rho}{2} \sum\limits_{k=1}^{\redN} x_{k}^{2}\right) \D x_{1} \ldots \D x_{\redN} \;.
\end{align}
In order to simplify the integral in \eqref{eqn:A1_Pe1_2}, we switch to the hyperspherical coordinate system \cite{book:weeks2002}, which is a generalization of the spherical coordinate system to higher dimensions. In this system, $\vect{x} \in \re^{\redN}$ is uniquely represented as $[r,\phii{1},\ldots,\phii{\redN-1}]$, where $r = \vectornorm{\vect{x}}$, and $\phi_{k}$ is the angle between $\vect{x}$ and the $k^{th}$ edge of $\set{D}_{f,q,i}$, $k=1,\ldots,\redN-1$. More precisely, let $\vect{v}_{_{k,q}}$ define the unit vector in the direction of the $k^{th}$ edge of $\set{D}_{f,q,i}$, for $k=1,\ldots,\redN$. Then, $\phi_{k} = \cos^{-1}(\vect{x}^{\rm T} \vect{v}_{k,q}/\vectornorm{\vect{x}})$, $k=1,\ldots,\redN-1$ (See Fig. \ref{fig:voronoi2} for a two-dimensional example.).

Next, we obtain the region of integration in \eqref{eqn:A1_Pe1_2} in hyperspherical coordinates. For any $\vect{x} \in \set{D}_{f,q,i} - \Omegaio$ represented by $[r,\phii{1},\ldots,\phii{\redN-1}]$, the parameter $r$  must satisfy 
\begin{align}
\label{eqn:r_limits}
\rbar \leq r \leq \infty \;,
\end{align}
where $\rbar$ is the distance of the point $\vect{x} \in \lbrace \vect{x} | \vect{a}_{f,i}^{\rm T}\vect{x} = b_{f,i} \rbrace$ from the origin. An expression for $\rbar$ can be found by representing the hyperplane $\vect{a}_{f,i}^{\rm T}\vect{x} = b_{f,i}$ in hyperspherical coordinates, using the inverse hyperspherical transform relations
\begin{align}
x_{k} &= r  \cos \phi_{k} \prod\limits_{k_{1}=1}^{k-1} \sin \phii{k_{1}}, k = 1,\ldots, \redN-1\;, \nonumber \\
\label{eqn:hyperspherical_cart}
x_{\redN} &= r \prod\limits_{k_{1}=1}^{\redN-1} \sin \phii{k_{1}}  \;,
\end{align}
and solving for $r$ as a function of $\boldsymbol{\phi}:=[\phi_{1},\ldots,\phi_{\redN-1}]$. Thus, we get  
\begin{align}
\label{eqn:rbar_def}
\rbar = \frac{b_{f,i}}{\sum\limits_{k =1}^{\redN-1} a_{k,f,i} \cos \phi_{k} \prod\limits_{k_{1}=1}^{k-1} \sin \phii{k_{1}} +a_{\redN-1 ,f,i}\prod\limits_{k_{1}=1}^{\redN-1} \sin \phii{k_{1}} } \;,
\end{align}
In \eqref{eqn:rbar_def}, $a_{k,f,i}$ is the $k^{th}$ element of $\vect{a}_{f,i}^{\rm T}$. Also, for any $\vect{x} \in \set{D}_{f,q,i} - \Omegaio$, it is seen that $\phi_{k}$ must be at least $0$ radians (if $\vect{x} = \alpha \vect{v}_{_{k,q}}, \alpha>0$), and at most $\fhi{k}$, which is the angle between $\vect{v}_{_{\redN,q}}$ and $\vect{v}_{_{k,q}}, k = 1,\ldots,\redN-1$. In other words,
\begin{align}
\label{eqn:A1_r_phi_limits}
&0 \leq \phii{k} \leq \cos^{-1}\left( \vect{v}_{_{\redN,q}}^{\rm T}\vect{v}_{_{k,q}}\right) =: \fhi{k} \;.
\end{align}
It is useful to note that $\fhi{k} \leq \pi$, since it is the angle between any two edges of the simplicial cone $\set{D}_{f,q,i}$, which is at most $\pi$.

Thus, \eqref{eqn:A1_Pe1_2} can now be reformulated in hyperspherical coordinates, with the limits of integration given by \eqref{eqn:r_limits} and \eqref{eqn:A1_r_phi_limits} as
\begin{align}
\label{eqn:A1_hyperspherical1}
\J = \left( \frac{\rho}{2 \pi} \right)^{\redN/2} \int\limits_{0}^{\fhi{\redN-1}}\ldots \int\limits_{0}^{\fhi{1}}\int\limits_{\rbar}^{\infty}r^{\redN-1} \sphi e^{-\rho r^{2}/2} \D r \D \phii{1}\ldots  \D \phii{\redN-1} \;,
\end{align}
where $\sphi := \prod_{k=1}^{\redN-2} \sin^{\redN-k-1} \phi_{k}$ arises from the Jacobian of the transformation. Substituting $u = r^{2}/2$ in \eqref{eqn:A1_hyperspherical1}, and changing the order of integration, we get
\begin{align}
\label{eqn:A1_Pe_bern1}
\Pei = \rho^{\redN/2}\int\limits_{0}^{\infty} e^{-\rho u} \tilde{\mu}_{i}(u) \D u \;,
\end{align}
where $\tilde{\mu}_{i}(u)$ is given by
\begin{align}
\label{eqn:A1_Pei_mui}
\tilde{\mu}_{i}(u):= \frac{1}{2 \pi^{\redN/2}} \dsum \sum\limits_{l=1}^{L} \int\limits_{\sheta{\redN-1}}^{\Sheta{\redN-1}}\ldots \int\limits_{\sheta{1}}^{\Sheta{1}} \sphi  u^{\redN/2 -1} I\left[\rbar^{2} \leq u \right]  \D \phii{1}\ldots \D \phii{\redN-1} \;.
\end{align}
In \eqref{eqn:A1_Pei_mui}, $L$ is the number of convex intervals of $[\phi_{1},\ldots,\phi_{\redN-1}]$ obtained after changing the order of integration, since the inverse function of $\rbar^{2}$ is not unique. We now show that $\tilde{\mu}_{i}(u) \geq 0$, which, together with Bernstein's Theorem implies that the integral in \eqref{eqn:A1_Pe_bern1} is equivalent to a {\cm} function of $\rho$. To this end, observe that the integrand in \eqref{eqn:A1_Pei_mui} is non-negative for $u \geq 0$, because $\sphi \geq 0$ for $\fhi{k} \in [0,\pi], k = 1,\ldots,\redN-1$. Consequently, the result obtained after the $\redN-1$ fold integration in \eqref{eqn:A1_Pei_mui} is also non-negative. Thus, $\tilde{\mu}_{i}(u)$, which is a scaled version of a sum of non-negative integrals, is also non-negative. Therefore, through Bernstein's Theorem, we can assert that $\Pei = \rho^{\redN/2} \ftildei$, where $\ftildei$ is a {\cm} function. 

Now, using \eqref{eqn:A1_Pe_bern1} in \eqref{eqn:A1_Pei_Pe}, we get
\begin{align}
\label{eqn:A1_AvgPe_1}
\AvgPe = \rho^{\redN/2} \int\limits_{0}^{\infty} e^{-\rho u} \tilde{\mu}(u) \D u \;,
\end{align}
where $\tilde{\mu}(u) := \sum_{i=1}^{M} \Pr[\vect{s} = \vect{s}_{i}] \tilde{\mu}_{i}(u)$. Thus, $\AvgPe = \rho^{\redN/2}\tilde{f}_{\rm cm}(\rho)$, where $\tilde{f}_{\rm cm}(\rho)$ is {\cm} through Bernstein's Theorem, because $\tilde{\mu}(u) \geq 0$ as it is a positive linear combination of non-negative functions $\mu_{i}(u), i = 1,\ldots,M$. 

Next, we strengthen the representation \eqref{eqn:A1_AvgPe_1} by showing that $\rho^{\alpha}\tilde{f}_{\rm cm}(\rho)$ is {\cm} for $\alpha = 1$. To this end, recall from Section \ref{subsec:math_cm} that the necessary and sufficient condition for a {\cm} function to be {\cm} of order $1$ is that its representing function be nonnegative and increasing. This is indeed the case for $\tilde{\mu}(u)$. Thus, we have just showed that $\tilde{f}_{\rm cm}(\rho)$ is {\cm} of order $\alpha=1$. Denoting $\fcm := \rho \tilde{f}_{\rm cm}(\rho)$, which we have just showed to be {\cm}, we get a stronger representation for the SER using \eqref{eqn:A1_AvgPe_1} as follows
\begin{align}
\label{eqn:A1_subfinal}
\AvgPe = \rho^{ \frac{\redN}{2}-1 } \fcm \;,
\end{align}
where $\fcm$ is {\cm}. 

Next, the support of the representing function of $\fcm$ is investigated. Let $\mu(u)$ be the representing function of $\fcm$. Accordingly, by applying integration by parts on \eqref{eqn:A1_AvgPe_1}, it is seen that 
\begin{align}
\mu(u) = \sum_{i=1}^{M}   \dsum \sum\limits_{l=1}^{L} \int\limits_{\sheta{\redN-1}}^{\Sheta{\redN-1}}\ldots \int\limits_{\sheta{1}}^{\Sheta{1}} & \sphi \frac{\Pr[\vect{s} = \vect{s}_{i}]}{2 \pi^{\redN/2}}  \left((\redN/2 -1) u^{\frac{\redN}{2} -2} I\left[\rbar^{2} \leq u \right] + \right. \nonumber \\ &\left. u^{\frac{\redN}{2}-1}I\left[\rbar^{2} = u \right] \right)\D \phii{1}\ldots \D \phii{\redN-1} \;,
\end{align} 
which is zero if $u <  \underset{q,f,i}{\min} \; \underset{\boldsymbol{\phi}}{\inf} \;\rbar^{2}$, and non-negative otherwise. Recalling the expression for $\rbar$ from \eqref{eqn:rbar_def}, it is immediately seen that $\underset{q,f,i}{\min} \; \underset{\boldsymbol{\phi}}{\inf}\;\rbar^{2} = \min_{f,i} b_{f,i}^{2}$. Observing that $\min_{f,i} b_{f,i}^{2} = d_{\rm min}^{2}/4$, we conclude that the support of $\mu$ is contained in $[d_{\rm min}^{2}/4, \infty)$.

This concludes the proof of the Theorem.
\section{Proof of Corollary \ref{thm:cm_BER_AWGN}}
\label{appendix:corr_1}
The direct part of the corollary for constellations with $\redN =1$ is immediate, since $\AvgPe$ is a positive linear combination of functions of the form $\Q{}{\sqrt{2 \rho \eta}}, \eta >0$, which is known to be {\cm} (see e.g. \cite{paper:STpaper11}). Also, the complete monotonicity of the SER for the case of $\redN =2$ is straightforward from Theorem \ref{thm:n_dim_error_rate}. To see the converse, we use the necessary and sufficient condition for a function $f(\rho)$ to be {\cm} of order $\alpha$, as described in Section \ref{subsec:math_cm}. For the case when $\redN$ is even, we assume $f(\rho) = f_{\rm cm}(\rho)$, where $f_{\rm cm}$ is as defined in Theorem \ref{thm:n_dim_error_rate}, and $\alpha = \redN/2 -1$. For the case of odd $\redN$, we assume $f(\rho) = \rho^{-1/2}f_{\rm cm}(\rho)$, and $\alpha = \lceil \redN/2 -1 \rceil$. The converse thus easily follows.
\section{Proof of Theorem \ref{thm:high_dim_cvx}}
\label{appendix:high_dim_cvx}
Recall from Theorem \ref{thm:n_dim_error_rate} that for a constellation with reduced dimension $\redN$, the SER can be written as 
\begin{align}
\label{eqn:A3_Pe}
\AvgPe = \rho^{p} \int\limits_{0}^{\infty}\exp(-\rho u) \mu(u) \D u \;,
\end{align}
where $p=  \redN/2-1 $, with $\mu(u)$ being zero when $u \in [0,d_{{\rm min}}^{2}/4)$, and non-negative otherwise. We now obtain sufficient conditions for $\AvgPedd \geq 0$. Differentiating \eqref{eqn:A3_Pe} twice with respect to $\rho$, we obtain
\begin{align}
\label{eqn:AvgPedd1}
\AvgPedd = \rho^{p-2}\int\limits_{0}^{\infty}e^{-\rho u} \mu(u) \left(u- \frac{p-\sqrt{p }}{\rho} \right)\left(u- \frac{p + \sqrt{p }}{\rho} \right) \D u \;,
\end{align}
where the differentiation under the integral sign in \eqref{eqn:A3_Pe} is permitted, because the limits of integration are independent of $\rho$. A sufficient condition for \eqref{eqn:AvgPedd1} to be non-negative is that the integrand in \eqref{eqn:AvgPedd1} is non-negative. Accordingly, observe that the integrand in \eqref{eqn:AvgPedd1} is non-negative when $u \geq (p + \sqrt{p})/\rho$. Thus, if the support of $\mu$ is a subset of $[(p + \sqrt{p})/\rho, \infty)$, it follows that $\AvgPedd \geq 0$. In other words, if $\rho \geq 4(p+ \sqrt{p})/d_{{\rm min}}^{2}$, we have $\AvgPedd \geq 0$, which proves the result.

\section{Proof of Theorem \ref{thm:high_ord_st_ord}}
\label{app:appendix_C}
For any two non-negative RVs $X_{1}$ and $X_{2}$, we have 
\begin{align}
\label{eqn_app_c_0}
X_{1} \leq_{\F_{p}} X_{2} \Leftrightarrow \E{}{X_{1}^{p} \exp(-\rho X_{1})} \geq \E{}{X_{2}^{p} \exp(-\rho X_{2})}, \forall \rho \geq 0.
\end{align}

We now establish the following:
\begin{align}
\label{eqn:app_c_2}
X_{1} \leq_{\F_{p}} X_{2} \Leftrightarrow \E{}{X_{1}^{p}f_{\rm cm}(\rho X_{1})} \geq \E{}{X_{1}^{p}f_{\rm cm}(\rho X_{1})} \;\forall \rho \geq 0, 
\end{align}
where $f_{\rm cm}(\rho):= \int_{0}^{\infty} \exp(-\rho u)\mu(u) \D u$ is a {\cm} function, for some $\mu(u) \geq 0$. Assume $u>0$, then $X_{1} \leq_{\F_{p}} X_{2} \Rightarrow \E{}{X_{1}^{p} \exp(-\rho X_{1} u)} \geq \E{}{X_{2}^{p} \exp(-\rho X_{2} u)} \; \forall \rho>0$. Next, observe that
\begin{align}
\label{eqn:app_c_3}
 \E{}{X_{1}^{p}f_{\rm cm}(\rho X_{1})} &= \E{}{\int\limits_{0}^{\infty} X_{1}^{p} \exp(-u \rho X_{1}) \mu(u) \D u } = \int\limits_{0}^{\infty} \E{}{X_{1}^{p} f_{\rm cm}(\rho X_{1})} \mu(u) \D u\\
 & \geq \int\limits_{0}^{\infty} \E{}{X_{2}^{p} f_{\rm cm}(\rho X_{2})} \mu(u) \D u =  \E{}{X_{2}^{p}f_{\rm cm}(\rho X_{2})},
\end{align}
$\forall \rho \geq 0$, provided the expectations exist. This proves the direct part of the theorem. To see the converse, let $f_{\rm cm}(\rho) = \exp(-\rho x)$, which is {\cm} in $\rho$ for each $x>0$.

\section{Proof of Theorem \ref{thm:high_ord_st_ord_neg}}
\label{app:high_ord_st_ord_neg}
Let $X_{1},\; X_{2}$ be non-negative RVs with PDFs $\pdf{X_{1}}{x}$ and $\pdf{X_{2}}{x}$ respectively. We now show that $X_{1} \orderl{\F_{p}} X_{2}$ cannot hold for all $p \geq 0$, by showing that $X_{1} \orderl{\F_{p}} X_{2}$ does not hold for every $p$ in a subset or $\re \cup \lbrace 0 \rbrace$, i.e., for $p \in \N \cup \lbrace 0 \rbrace$. In order to satisfy Theorem \ref{thm:high_ord_st_ord} for every $p \in \set{N} \cup \lbrace 0 \rbrace$, we require for every $p$
\begin{align}
\label{eqn:app4_e1}
(-1)^{p} \frac{d^{p}}{d \rho^{p}}\left[ \LT{}{\rho}{X_{1}} - \LT{}{\rho}{X_{2}}\right] \geq 0 \; \forall \rho ,
\end{align}
where we have used the identity $\int_{0}^{\infty}s^{p} \exp(-\rho s) \mu(s) \D s =(-1)^{p}(\partial^{p}/\partial \rho^{p})\int_{0}^{\infty} \exp(-\rho s) \mu(s) \D s$. Recalling the definition of a {\cm} function from \eqref{eqn:cm_def}, we gather that $\LT{}{\rho}{X_{1}} - \LT{}{\rho}{X_{2}}$ in \eqref{eqn:app4_e1} must be a {\cm} function. Consequently, Bernstein's Theorem mandates that $\pdf{X_{1}}{x} - \pdf{X_{2}}{x} \geq 0 \; \forall x$. However, this condition is never satisfied by any pair of random variables, since both the density functions must individually integrate to unity, which cannot be the case if $\pdf{X_{1}}{x} \geq \pdf{X_{2}}{x} \; \forall x$. Thus, the Theorem follows.

\bibliographystyle{IEEEtran}
\nocite{*}
\bibliography{references}
\newpage
\newpage
\begin{figure}
\begin{center}
\includegraphics[width=11cm,keepaspectratio]{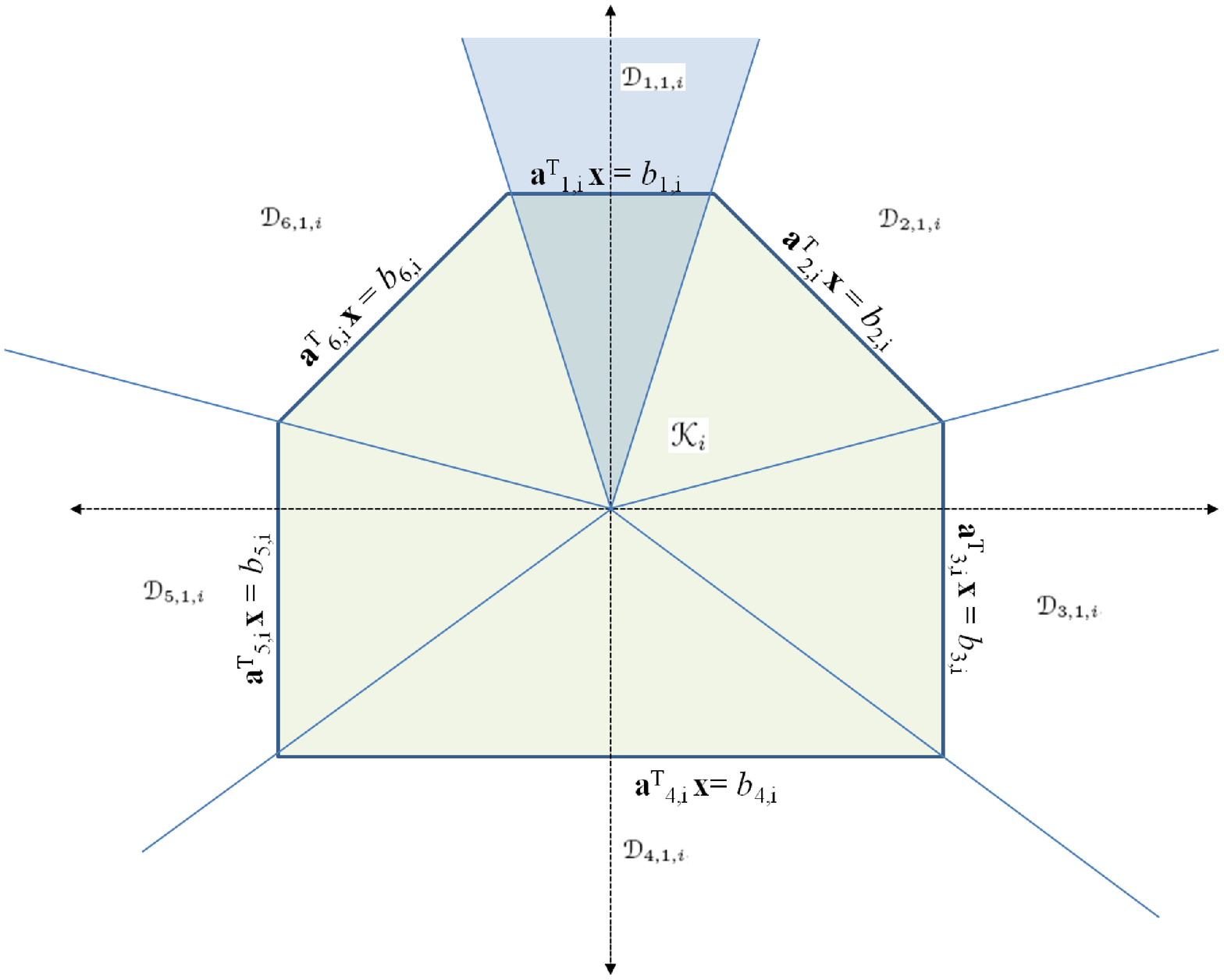}
\caption{Voronoi region for signal point $\vect{s}_{i}^{*} \in \set{S}^{*}$, where the reduced dimension of the constellation $\set{S}$ is $\redN =2$. The origin of the coordinate axes is shifted to $\vect{s}_{i}^{*}$. Using Lemma 1, $\re^{N^{*}}$ is decomposed into a collection of $2$-dimensional simplicial cones $\set{X}_{i}:= \lbrace \set{D}_{1,1,i}, \ldots,\set{D}_{6,1,i} \rbrace$ using the facets of $\set{K}_{i}$.} \label{fig:voronoi1}
\end{center}
\end{figure}

\begin{figure}
\begin{center}
\includegraphics[width=6cm,keepaspectratio]{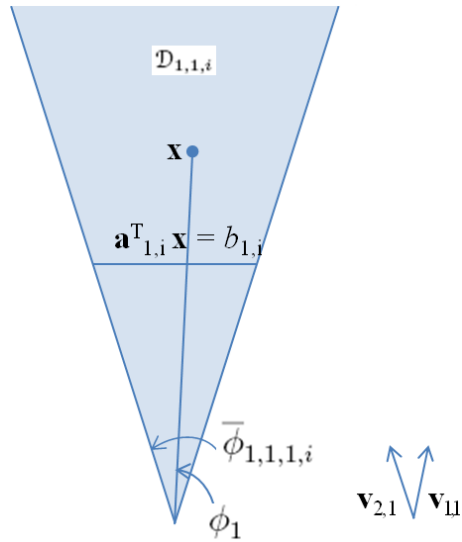}
\caption{The $2$-dimensional simplicial cone $\set{D}_{1,1,i}$ obtained using Lemma 1 with $\set{K}_{i}$, represented in hyperspherical coordinates.}
\label{fig:voronoi2}
\end{center}
\end{figure}

\begin{figure}
\begin{center}
\includegraphics[width=9cm,keepaspectratio]{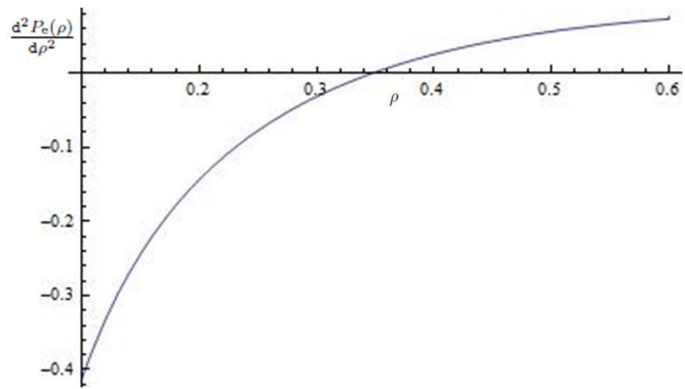}
\caption{Second derivative of $\AvgPe$ for the $3$-dimensional square QAM constellation. Since the SER is not convex, this is an example of a constellation with $\redN =3$ whose SER is not {\cm}.}
\label{fig:QAM3D}
\end{center}
\end{figure}

\end{document}